\documentstyle[11pt,baaa-eng,twoside,epsf,psfig]{article}
\begin{document}
\vskip 1.0cm
\markboth{S.~E. Nuza et al.}{GRBs host galaxies}
\pagestyle{myheadings}

\vspace*{0.5cm}
\parindent 0pt{ COMUNICACI\'ON DE TRABAJO -- CONTRIBUTED PAPER } 
\vskip 0.3cm
\title{Study of the relationship between the gamma ray bursts and their host 
galaxies} 

\author{Sebasti\'an E. Nuza}
\affil{Instituto de Astronom\'{\i}a y F\'{\i}sica del Espacio, Buenos Aires, Argentina; sebasn@iafe.uba.ar}

\author{Patricia B. Tissera}
\affil{Instituto de Astronom\'{\i}a y F\'{\i}sica del Espacio, Buenos Aires, Argentina; patricia@iafe.uba.ar}

\author{Leonardo J. Pellizza}
\affil{Service d'Astrophysique, DSM/DAPNIA, CEA, Saclay, France; 
leonardo.pellizza@cea.fr}

\author{Diego G. Lambas}
\affil{IATE, Observatorio Astron\'omico de C\'ordoba, C\'ordoba, Argentina; dgl@oac.uncor.edu}

\author{Cecilia Scannapieco}
\affil{Instituto de Astronom\'{\i}a y F\'{\i}sica del Espacio, Buenos Aires, Argentina; cecilia@iafe.uba.ar}

\author{Mar\'{\i}a E. De Rossi}
\affil{Instituto de Astronom\'{\i}a y F\'{\i}sica del Espacio, Buenos Aires, Argentina; derossi@iafe.uba.ar}

\begin{abstract} 
Gamma ray bursts (GRBs) belong to the most energetic events in the
Universe. Recently, the extragalactic nature of these sources has been 
confirmed with the discovery of several host galaxies (HGs) and the 
measurement of their redshifts. 
To explain the origin of GRBs various models have been proposed, 
among which the coalescence of compact objects and the {\it collapsar} 
scenarios are the most representative, being the collapsar 
model one of the most accepted to explain the long duration GRBs.
A natural consequence of this model is that the GRBs would trace the 
star formation rate (SFR) of their HGs. 
In this contributed paper we present preliminary results of the 
development of a Montecarlo-based code for collapsar event formation which 
is coupled to chemical-cosmological simulations aiming at studying the 
properties of HGs in a hierarchical scenario.
\end{abstract}

\begin{resumen}  
Las explosiones de radiaci\'on gamma (ERG) se encuentran entre los eventos 
m\'as energ\'eticos del Universo. Recientemente, la naturaleza 
extragal\'actica de estas fuentes fue confirmada con el descubrimiento de 
varias de sus galaxias hu\'esped y la medici\'on de sus corrimientos 
al rojo. Para explicar su origen, varios modelos han sido propuestos, 
entre los cuales las colisiones de objetos compactos en sistemas binarios y 
los llamados {\it collapsars} son los m\'as representativos, siendo 
este \'ultimo modelo uno de los m\'as aceptados para explicar las ERG de 
larga duraci\'on. 
Una consecuencia natural del mismo es que las ERG trazar\'{\i}an la 
tasa de formaci\'on estelar de sus respectivas galaxias. 
En esta comunicaci\'on se presentan resultados preliminares del desarrollo 
de un c\'odigo Montecarlo para la formaci\'on de collapsars en 
simulaciones qu\'{\i}mico-cosmol\'ogicas, con el fin de estudiar las 
propiedades de sus galaxias hu\'esped en un escenario jer\'arquico.
\end{resumen}

\section{Introduction}

The gamma ray bursts (GRBs) are the most energetic electromagnetic events in 
the Universe (e.g. Piran et al. 2000). One of the preferred models to explain 
the long duration (i.e., $> 2$ s) GRBs is the star core collapse into a 
black hole produced in a supernova type SNIb/c explosion 
(Mac Fayden, Woosley \& Heger 2001). This mechanism of GRB formation is also 
known as the {\it collapsar} scenario. 
This model is linked to the evolution of massive stars which normally have 
a mean lifetime of several million years implying that the typical lifetime 
of a GRB progenitor system is negligible in cosmological terms. 
So, it turns to be natural to consider these events as possible tracers of 
the cosmic star formation history up to high redshifts. 
In particular, the GRBs would permit to obtain information about star 
formation regions in galactic systems with different levels of evolution.

Recently, Courty et al. (2004) made use of structure formation 
simulations in order to identify galactic populations in the 
simulated sample capable to reproduce the observational features of the 
observed  HGs. These galaxies show a trend to be bluer and sub-luminous 
(e.g. Le Floc'h et al. 2003). Their analysis seems to confirm the connection 
with the star formation rate (SFR) if the GRBs events are formed in galaxies 
with high star formation efficiency. In this work we develop a GRB event 
generator based on the collapsar model for the progenitors and study the 
properties of the HGs in cosmological simulations.

\begin{figure}  
\hbox{ 
   \psfig{figure=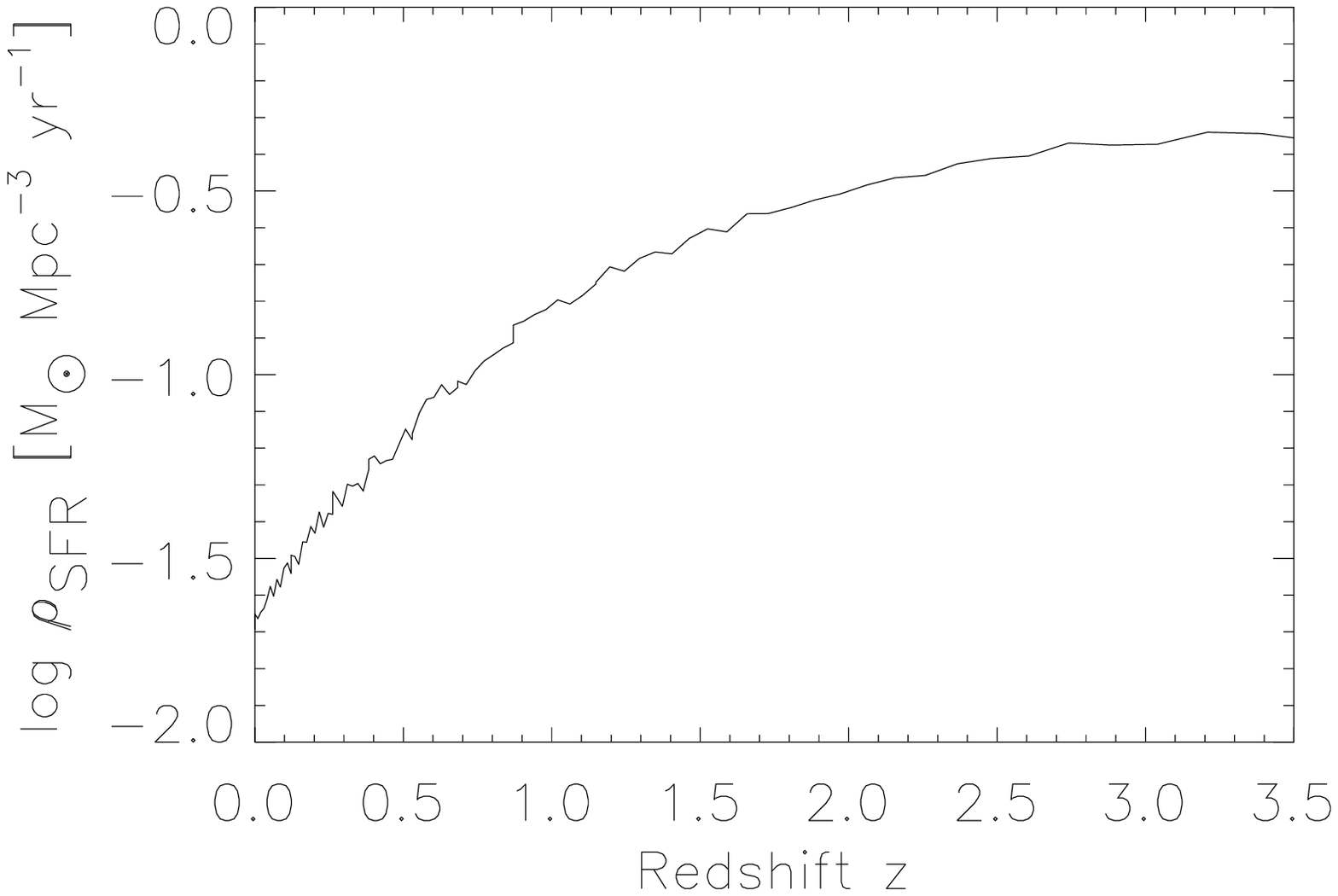,width=6.cm}
          \hspace*{0.2cm}
   \psfig{figure=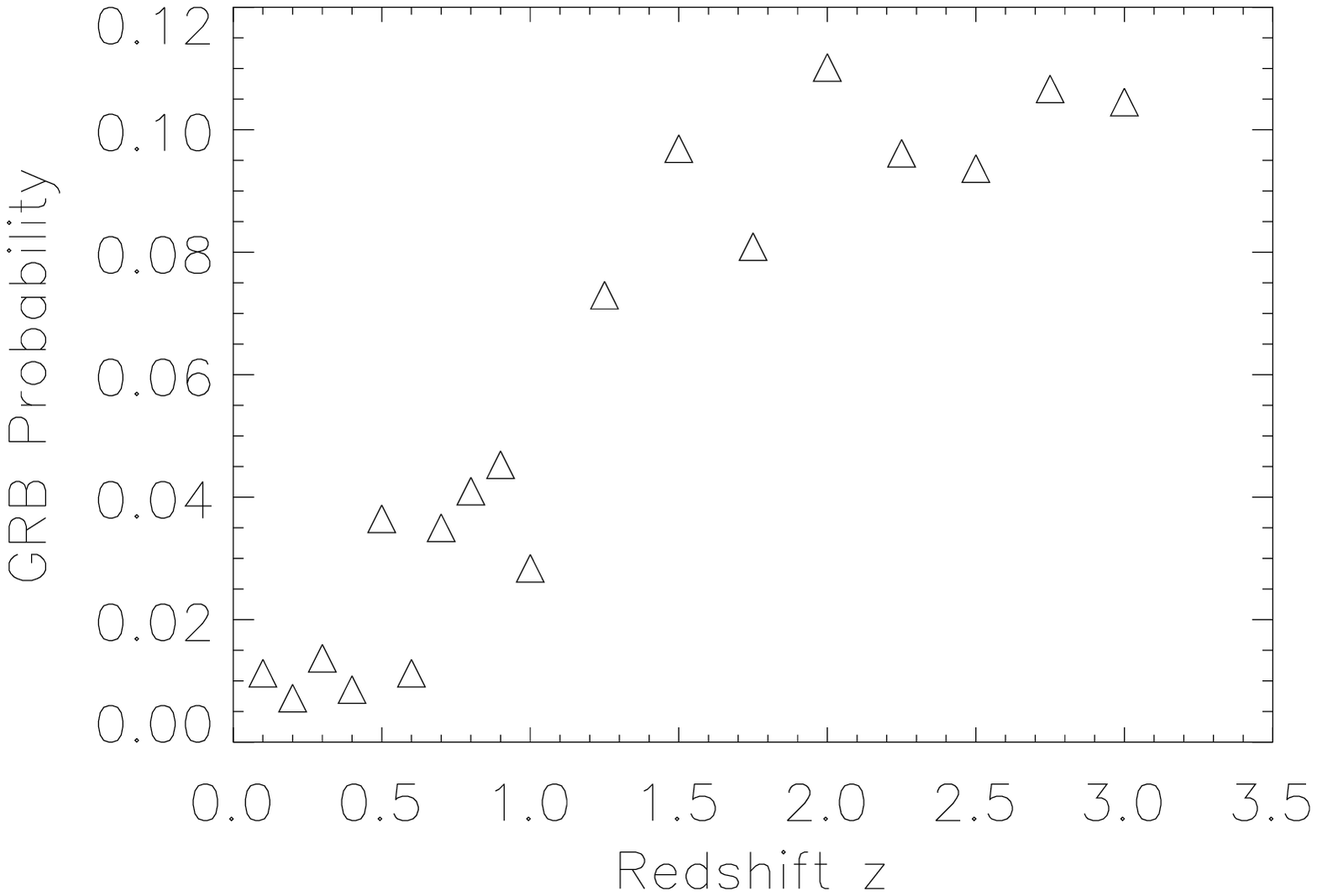,width=6.cm}
        }
\caption{{\it Left:}~~Cosmic SFR density for the simulated box of 10 Mpc 
$h^{-1}$ comoving size as a function of redshift. 
{\it Right:}~~Simulated~{\it probability}~for GRBs occurrence as a function 
of redshift.}
\end{figure}
\vskip 1.1cm

\section {GRB Montecarlo code and discussion}

The GRB event algorithm has been designed to work coupled to numerical 
simulations in a cosmological framework. We used the simulated galaxy 
catalog constructed by De Rossi et al. (2005) from simulations performed with 
the chemo-dynamical code of Scannapieco et al. (2005) developed within 
{\small{GADGET-2}} (Springel \& Hernquist 2002).
The gas component is followed using the Smoothed Particle Hydrodynamics 
technique while the dark matter component is followed using a Tree-Particle 
Mesh method.
This chemo-dynamical code describes the enrichment of the interstellar 
medium by SNII and SNIa supernova explosions. 
The cosmological model assumed is defined by the following 
set of cosmological parameters: $\Omega_{\rm M}=0.3$, $\Omega_{\Lambda}=0.7$, 
$\Omega_{\rm b}=0.04$ and $H_0 = 100$ $h$ km s$^{-1}$ Mpc$^{-1}$ with $h=0.7$. 
The simulations represent a typical region of 10 Mpc $h^{-1}$ comoving size 
with $2 \times 80^3$ particles which translates into an initial  mass 
resolution  of $2 \times 10^7~{\rm M_{\odot}}$ $h^{-1}$ and 
$2 \times 10^8~{\rm M_{\odot}}$ $h^{-1}$ for the  gas and dark matter 
particles, respectively. The catalog of simulated galaxies provides 
information on the gas, stellar and dark matter components for the 
redshift range $z = [0, 3]$.
A Montecarlo-based code was developed to generate GRBs events in each 
simulated galaxy of the catalog. We assumed Poissonian statistics to emulate 
the probability distribution of the GRBs events. The GRB generator selects 
young stars as possible candidates, where SNIb/c events may occur, and 
consequently, collapsar events can develope. 
For that, we adopted a cut-off in stellar ages of $t_{\rm c}= 10^7$ yr. 
A median rate of collapsar events ($R_{\rm coll}$) consistent 
with that estimated by Fryer et al. (1999), 
$R_{\rm coll} \sim 10-1000$ Myr$^{-1}$ Galaxy$^{-1}$ where Galaxy represents 
a typical galaxy, was assumed.
In particular, in this work we used $R_{\rm coll}=100$ Myr$^{-1}$ 
Galaxy$^{-1}$ and a typical galaxy mass of $10^{11}~{\rm M_{\odot}}$. 
A total of 500 Montecarlo realizations was perfomed for every selected stellar 
particle in each simulated galaxy  in  the catalog. Our preliminary results 
can be seen in Figures 1 and 2.

Figure 1 shows the simulated comoving SFR density 
($\rho_{\rm SFR}$, left panel) and the simulated {\it probability} of 
GRB event occurrence (right panel) as a function of redshift.
As it can be seen from Figure 1, assuming the collapsar scenario, where GRBs 
originate from massive stars, produces the expected behaviour of GRBs being 
good tracers of the cosmic SFR history.
Figure 2 shows the SFR efficiency ($\epsilon_{\rm SFR}$, defined as the SFR 
normalized to the total stellar mass at each analysed redshift) versus 
circular velocities ($V_{\rm opt}$) of the HGs from $z=3$.
It can be seen from this figure that the SFR efficiency anticorrelates with 
$V_{\rm opt}$, so that the higher the efficiency, the smaller the systems 
(i.e. slower rotating systems). 
Note that, at all analyzed $z$, more than 50 per cent of the simulated HGs 
have $V_{\rm opt} < 100~{\rm km~s^{-1}}$. According to 
the Tully-Fisher relation, this result implies that high SFR efficiencies are 
associated to sub-luminous systems in agreement with observations. Note
also that the SFR efficiency increases with increasing redshifts and that for 
$z > 1$, all HGs seem to have similar values.

Interesting improvements for the future would be to include an analysis 
based on the simulated colours and metallicities of the simulated HGs. 
Numerical simulations with higher resolution are also being analysed in 
order to confirm these trends.   

\begin{figure}
{\hspace*{1.5cm}  \psfig{figure=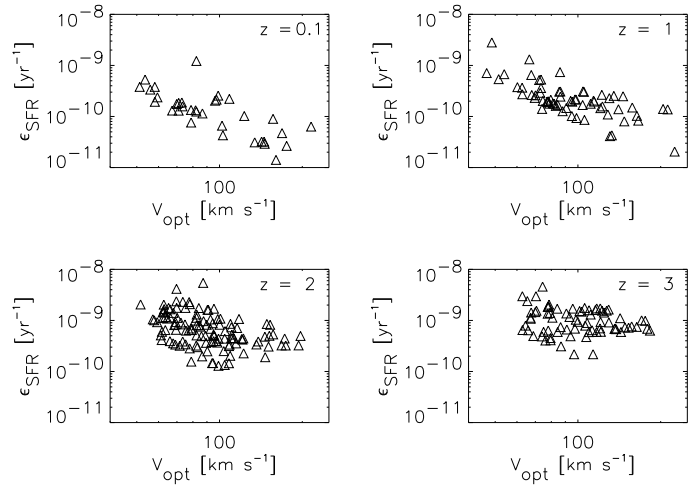,width=9.cm}}
\caption{SFR efficiency versus circular velocity for the simulated HGs sample 
from $z=3$ to $z=0.1$.}
\end{figure}

\acknowledgments 
Numerical simulations were run on Ingeld and HOPE PC-clusters at IAFE. 
We acknowledge financial support from CONICET, ANPCyT, Fundaci\'on Antorchas, 
SECyT-UNC and LENAC network. We also thank the referee for several comments 
that helped to improve this contributed paper.


\begin{references}
\reference Courty, S., Bj$\ddot{\rm o}$rnsson, G. \& Gudmundsson, E. H., 2004, 
MNRAS, 354, 581
\reference De Rossi, M. E., Tissera, P. B. \& Scannapieco, C., 2005, 
Ap\&SS, 329, 15 
\reference Fryer, C., Woosley, S. E. \& Hartmann, D., 1999, ApJ, 526, 152 
\reference Le Floc'h, E., Duc, P. A., Mirabel, I. F., et al., 2003, 
A\&A, 400, 499
\reference MacFadyen, A. I., Woosley, S. E. \&  Heger, A., 2001, ApJ, 550, 410
\reference Piran, T., 2000, Physics Reports, 333, 529
\reference Scannapieco C., Tissera P. B., White S. D. M. \& Springel V., 2005, 
MNRAS, 364, 552
\reference Springel V. \& Hernquist L., 2002, MNRAS, 333, 649
\end{references}
\end{document}